\documentclass[english]{lni}
\usepackage[nolist]{acronym}
\usepackage{booktabs}
\usepackage{pifont}
\usepackage{subcaption}
\usepackage{tkz-kiviat}
\usepackage{adjustbox}
\usepackage[dvipsnames]{xcolor}
\usepackage{float}
\usetikzlibrary{arrows}
\begin{document}
\title{Investigating Retargetability Claims for Quantum Compilers}
\author[1]{Luke Southall}{luke.southall@student.kit.edu}{0009-0007-3659-0830}
\author[1]{Joshua Ammermann}{joshua.ammermann@kit.edu}{0000-0001-5533-7274}
\author[1]{Rinor Kelmendi}{rinor.kelmendi@kit.edu}{0009-0003-0951-6871}
\author[1]{Domenik Eichhorn}{domenik.eichhorn@kit.edu}{0000-0001-9428-024X}
\author[1]{Ina Schaefer}{ina.schaefer@kit.edu}{0000-0002-7153-761X}
\affil[1]{Karlsruhe Institute of Technology} 
\maketitle

\begin{abstract}
In the NISQ-era, there is a wide variety of hardware manufacturers building quantum computers. Each of these companies may choose different approaches and hardware architectures for their machines. This poses a problem for quantum software engineering, as the retargetability of quantum programs across different hardware platforms becomes a non-trivial challenge. In response to this problem, various retargetable quantum compilers have been presented in the scientific literature. These promise the ability to compile software for different hardware platforms, enabling retargetability for quantum software. 
In this paper, we develop and apply a metric by which the retargetability of the quantum compilers can be assessed. We develop and run a study to analyze key aspects regarding the retargetability of the compilers Tket, Qiskit, and ProjectQ. Our findings indicate that Tket demonstrates the highest level of retargetability, closely followed by Qiskit, while ProjectQ lags behind. These results provide insights for quantum software developers in selecting appropriate compilers for their use-cases, and highlight areas for improvement in quantum compilers.
\end{abstract}

\begin{keywords}
Quantum Computing \and Quantum Compiler \and Retargetability \and User Study
\end{keywords}

\section{Introduction}

Quantum computing is a rapidly evolving field that draws a lot of attention from research and industry due to the potential to use quantum mechanical effects to solve specific tasks faster than classical computers~\cite{national2018quantum}.
Currently, we are in the so-called Noisy Intermediate-Scale Quantum (NISQ) era, giving us access to devices,
which, apart from having various problems like high error rates and short coherence times, have a limited number of qubits. 
At present, these hardware limitations prevent a demonstrable advantage of quantum algorithms for relevant problems~\cite{preskill2018quantum}.
Furthermore, quantum hardware varies significantly between different manufacturers, as there are different ways to physically implement a quantum information unit~\cite{compreviewqiu}. 

The heterogeneity of quantum computing hardware leads to many problems in quantum software engineering, in particular with respect to cross-platform compatibility~\cite{murilloChallengesQuantumSoftware2024}. 
Given this limitation, executing programs on different hardware becomes a non-trivial challenge, akin to the challenges in classical computing when compiling programs to run on different processor architectures like ARM or x86. The important distinction between this challenge in the classical space and the challenge in the quantum space is the lack of standardized architectures in the quantum computing field. The x86 and ARM architectures are ubiquitous, whereas in the quantum space there are a significant number of different hardware constraints that vary in many aspects, such as environmental requirements, qubit connectivity and topology, scalability, and coherence times. This creates overhead for developers, who may have to rewrite their software to accommodate different hardware constraints, and thereby hinders progress in quantum software development.
As a response to the high degree of heterogeneous quantum technologies, retargetable quantum compilers translate code in a given language to code specific to a changing quantum architecture supported by the compiler.

In this paper, we will explore the field of quantum compilers that potentially enable retargetability, that is, they are able to work with multiple hardware architectures and input languages with full cross-compatibility. This paper makes two primary contributions: 
\begin{enumerate}
    \item We develop a methodology to measure the retargetability of a quantum compiler. Here, (a) we  identify important dimensions of retargetability and use these to design a retargetability metric, and (b) we design a user study in which participants evaluate the retargetability of quantum compilers based on backends they implement. 
    \item We apply this methodology to empirically assess three popular quantum compilers regarding retargetability: Tket, Qiskit, and ProjectQ. We apply our developed metric, resulting in a retargetability score for each of the compilers. We conduct the planned study with six participants, which implement a backend for each compiler and then rate identified aspects for retargetability. By combination we provide a comparative analysis of retargetability for the chosen quantum compilers.
    We find that Tket has the highest retargetability followed closely by Qiksit. ProjectQ receives the lowest score for retargetability.
\end{enumerate}

\section{Background and Related Work}

Quantum compilers bridge the gap between quantum algorithms and physical quantum hardware, translating high-level quantum programs into low-level quantum circuits. 
Like classical compilers, they use an intermediate representation generated from high-level code to perform hardware-independent and hardware-dependent optimizations before translation to hardware-specific representations. 
Well-known examples are compilers from established quantum programming languages such as Cirq~\cite{GOOGLE:CIRQ}, Qiskit~\cite{IBM:QISKIT}, Q\#~\cite{Svore_2018}, and Qrisp~\cite{seidel2024qrisp}; language-independent compilers such as Tket~\cite{Sivarajah_2020} and ProjectQ~\cite{Steiger_2018}; and specialized tools for circuit optimization such as PyZX~\cite{kissinger2020Pyzx} and staq~\cite{amy2020staq}. 
Given the broad range of compilers, several studies have been conducted to evaluate their performance. 
Arline~\cite{kharkov2022arline}, Benchpress~\cite{nation2024benchmarking}, MQT-Bench~\cite{quetschlich2023mqt}, and Qasmbench~\cite{li2023qasmbench} are large-scale benchmarks that propose to study metrics such as qubit numbers, the number of specific gates, and the impact of different transpilers / transpiler passes / optimization levels to a circuit. 
More specialized works use similar metrics to directly study the impact of intermediate representations~\cite{meijer2025comparison}, or the impact of transpilation pass combinations~\cite{dangwal2024compass}.

A metric that is usually not covered in these studies is a compiler's retargetability, which describes the compilers ability to generate valid instructions (circuits) for different hardware backends.
Retargetability, remains complex in quantum compilation due to diverse hardware architectures. Each platform has unique gate sets, qubit connectivity, and error characteristics, making universally adaptable compilation challenging. While established compilers such as Qiskit and ProjectQ support the implementation of multiple backends, there are only two compiler frameworks that explicitly claim to be retargetable: Tket~\cite{Sivarajah_2020} and Weaver~\cite{kirmemics2025weaver}.
Although these contributions show progress in the direction of retargetability, a widespread adaption of retargetability requires further standardization of the quantum computing stack and enhanced compilation techniques.

We expect retargetable compilers to become an important aspect of future-proof quantum software stacks. 
Works that quantify the retargetability of quantum compilers are Gierisch~\cite{Gierisch:AnaEvalQC}, K\i rmemi\c{s}~et~al.~\cite{kirmemics2025weaver}, and Salm~et~al.~\cite{Salm2021_CompilerComparison}.
Gierisch~\cite{Gierisch:AnaEvalQC} implements an evaluation framework comparing compilation outputs, particularly one-qubit and two-qubit gate counts across optimization levels and target architectures, also measuring compilation time. K\i rmemi\c{s}~et~al. introduce Weaver~\cite{kirmemics2025weaver} as a verifiable retargetable compiler framework, specially designed for FPQA quantum architectures. To quantify retargetability, they propose three metrics:
(1)~extensiblilty,~(2)~performance~and~fidelity,~and~(3)~verifiability. 
Salm~et~al.~\cite{Salm2021_CompilerComparison} present a comparison of multiple quantum compilers regarding retargetability, listing supported frontends, backends, and hardware compatibility options.
Similarly to the methods that will be presented in this paper, they profile for selected retargetable compilers and compare these metrics with the goal of developing a tool that compiles quantum code across different compilers to the same target backend.

Given these recent contributions, we can observe that retargetability becomes an important feature for quantum compilation frameworks. Future compilers need to be able to adapt to multiple backends and technologies. 
However, we also see that there is currently no standardized way to assess if a compiler is actually retargetable.
In this paper, we make an initial contribution to tackle this problem and take inspiration from the existing evaluations on compiler performance and the initial three studies that directly look at retargetability.

\section{Methodology for Measuring Retargetability}
\label{section3}
To evaluate retargetability claims of quantum compilers, we first have to address \textbf{RQ1: What are important aspects of retargetability, and how can we measure them?}
To identify retargetability aspects, one can learn from successful classical compilers, but also has to consider specifics of the quantum software environment.
For classical compilers, modular and device-agnostic architectures, as well as flexible optimization pipelines, are important for retargetability. 
The lack of standardization and the rapidly changing quantum software environment are currently hindering better retargetability of quantum software. 
Finally, for both classical and quantum compilers, high quality, comprehensive APIs and documentation are necessary to achieve practical retargetability. 
A complete representation of important aspects for retargetability for quantum compilers has to address all these aspects.
To answer \textbf{RQ1} we propose a novel methodology to evaluate the retargetability of quantum compilers across five dimensions. Our method comprises two approaches: a metric-based assessment of retargetability across four dimensions, and as fifth dimension, a complementary practical user study where participants implement a backend.

\subsection{Metric-based Assessment of Retargetability}
We propose a metric that assesses the retargetability of quantum compilers across four dimensions.
Each of these dimensions is evaluated with a set of questions available at~\cite{zenodolink}, which are rated on a five-point Likert-scale, where higher scores indicate better performance. In the following paragraphs, we briefly explain each dimension and our reasoning behind it.

\textbf{Compilation Strategy Flexibility:} For better retargetability, it is important to have fine-grained control over compilation processes while remaining hardware independent. We rate the level of control over the compilation process, how configurable compiler passes are, how comprehensive options are to specify hardware-specific constraints, if there are presets for compilation strategies and if there is the possibility to create custom compilation passes.

\textbf{Standardization Compliance:} Currently missing uniform standards in quantum computing create compatibility issues, but adaptation of (quasi-) standards improves the interoperability of other quantum software and therefore increases retargetability. We rate support for the \ac{openqasm}, support for the \ac{qir}, support for additional interchangeable formats, active involvement in standardization efforts, and overall interoperability with other quantum software tools.

\textbf{Community and Ecosystem Integration:} Due to quantum computing being a rapidly evolving and changing field, it is important for retargetable compilers to have an active community and integrate well with hardware providers. In this dimension, we rate the active integration with providers, the growth in supported integrations, the ease of using the extension system, effectiveness of backend distribution, and the community engagement.

\textbf{Device-Agnostic Compiler Architecture:} Compilers should be designed without any coupling to specific manufacturer architectures to increase retargetability. We rate the level of modular design, hardware agnostic design, the adaptability of the optimization pipeline, the adaptability to different quantum gate sets, and the priority of hardware agnosticism in the design. Additionally, this dimension can be complimented by a Sonar Qube~\cite{sonarqube} evaluation of the code-bases software quality. Sonar Qube provides scores for the codebase in the categories security, reliability, and maintainability. 

\subsection{User Study for Documentation and API Quality}

To empirically evaluate the \textbf{Documentation and API Quality} dimension, we design a study to find evidence of practical retargetability from a developer perspective.
Developers have to solve an implementation task to create a new backend for a quantum compiler and then answer a questionnaire available at~\cite{zenodolink}. For each compiler, we identify potential hurdles based on the compiler's documentation and API. Participants are provided with hints they can use, designed to replicate the kind of guidance they might receive from community forums. The usefulness of these hints is tracked as part each compiler's questionnaire.

\textbf{Task Selection:}
We created a backend implementation task that requires participants to create a backend implementation for a quantum simulator.
In this task the participants need to gain a basic understanding of the simulator. It also requires the participants to understand the documentation and API. These two factors reveal Documentation and API quality issues.

\textbf{Study Protocol:}
Participants are provided with standardized instructions for each compiler, including the task description, links to relevant documentation sections, and basic setup instructions for the compiler and the simulator. To ensure fair comparison, instructions are designed to be as close to each other as possible for different compilers, providing the same level of initial guidance without prescribing specific implementation approaches.
In the final questionnaire, participants rate their experience across multiple subdimensions for Documentation and API quality. Backend documentation comprehensiveness measures whether the documentation provides sufficient information on how to implement a new backend. API clarity and intuitiveness assess how easily participants understand the backend API structure. Documentation organization and navigability evaluate how easily participants find relevant information. Example and tutorial quality measures the usefulness of the provided examples for understanding a backend implementation. Hint utility assesses whether provided hints are necessary, and if so, whether they effectively address common obstacles. Overall, implementation process difficulty captures participants' subjective assessment of the entire implementation experience.
Participants provide ratings for these aspects after completing the implementation for each compiler, allowing for a direct comparison based on recent experience.
We designed the questionnaire using best practices~\cite{Dillman2024}, evaluating questions on a five-point Likert-scale, where higher scores indicate better performance. 

\subsection{Overall Evaluation Approach}

The combination of the metric-based retargetability assessment and the study provides a complementary perspective on retargetability.
We combine the results of these evaluations to reach a final, combined retargetability score. The score is calculated for each dimension $i \in [1, 5]$ using the formula $s_{total} = \sum_{i=1}^{5} w_i \cdot s_i$, where $s_i \in [1, 5]$ is the categories score, $w_i \in [0, 1]$ is the categories weight and $\sum_{i=1}^{5} w_i = 1$. For this work we use an arithmetic mean with $w_{1,...,5}=0.2$, granting all evaluated dimensions the same importance. We propose that weights should be assigned on a case-by-case basis.
\section{Retargetability of Tket, Qiskit and ProjectQ}

Our second research question is \textbf{RQ2: How retargetable are existing quantum compilers according to our methodology?} 
To assess this, we apply our evaluation approach to three quantum compilers: \textit{Tket}~\cite{Sivarajah_2020}, \textit{Qiskit}~\cite{IBM:QISKIT} and \textit{ProjectQ}~\cite{Steiger_2018}. We chose Tket, as it is explicitly claimed to be a retargetable quantum compiler. Qiskit was chosen because it is very popular and offers a guide on how to implement custom backends. ProjectQ also does not explicitly claim retargetability. However, the compiler is designed with hardware-agnosticism in mind and offers the possibility to integrate custom backends. 
Another suitable compiler for would have been the Weaver compiler, which also directly claims to be retargetable. However, it was only released after our study was designed.

\subsection{Metric-based Assessment of Tket, Qiskit and ProjectQ}

Following, we assess the retargetability using the dimensions described in Section~\ref{section3}.

\textbf{Compilation Strategy Flexibility:}
The three compilers show varying flexibility, as seen in Table~\ref{tab:met-compilation-flexibility}. 
\textit{Tket} provides excellent flexibility, receiving the highest possible rating in all subcategories. It offers extensive user control over compilation pass selection, ordering, and customization. Pre-built compilation pass options are very extensive and allow for easy selection and configuration. Tket provides detailed options for specifying and customizing hardware constraints, including architecture options and predicates for checking gate sets and connectivity. Multiple pre-defined strategies are available for different use cases, both hardware-independent and hardware-specific. Custom compilation pass support is very robust through varied pass options, for example, \texttt{CustomPass} for function-based transformations.
\textit{Qiskit} also offers excellent flexibility. Users have extensive control over the transpilation process, including pass selection, ordering, and customization. Qiskit provides comprehensive options for selecting and configuring pre-built compilation passes, supported by tutorials. Hardware-specific constraint handling allows detailed specification and customization of hardware constraints. Qiskit offers multiple pre-defined compilation pipelines, especially for various IBM machines and architectures. The framework supports the creation and integration of custom compilation passes, also backed by tutorials. This allows users to adapt the compilation process to their specific needs and hardware constraints.
\textit{ProjectQ} offers more limited flexibility. Users have some control over compilation passes through \texttt{Engine} lists with full control over ordering and customization, but limited pre-built selection options. The pre-build compilation pass options are rated as neutral with compilation engines for optimization, routing, and translation, but missing engines for other functionality, e.g. synthesis engines. The hardware constraint specification is somewhat limited, supporting connectivity and gate sets, but lacking gate fidelities and noise models. Pre-defined compilation strategies are available for several providers, including AQT, IBM, and IonQ. Custom pass support is very robust as ProjectQ supports the creation and integration of custom compilation passes through its \texttt{Engine} system.

\begin{table}[tbh]
\centering
\scriptsize
\begin{tabular}{p{8.8cm}|c|c|c}
\toprule
\textbf{Question} & \textbf{Tket} & \textbf{Qiskit} & \textbf{ProjectQ} \\
\midrule
How much or how little control does a user have over compilation passes (selection, ordering, customization)? & 5 & 5 & 3 \\
How extensive or limited are the pre-built compilation passes available for selection and configuration? & 5 & 5 & 3 \\
How comprehensive or limited are the options for specifying and customizing hardware-specific constraints (e.g., qubit connectivity, gate fidelities)? & 5 & 5 & 2 \\
How extensive or limited are the preset compilation strategies available (e.g., noise mitigation, depth reduction)? & 5 & 5 & 2 \\
How robust or limited is the support for creating and integrating custom compilation passes? & 5 & 5 & 5 \\
\midrule
\textbf{Average} & 5 & 5 & 3 \\
\bottomrule
\end{tabular}
\caption{Questions and results for the Compilation Strategy Flexibility dimension.}
\label{tab:met-compilation-flexibility}
\end{table}

\textbf{Standardization Compliance:}
As shown in Table~\ref{tab:met-standardization}, \textit{Tket} demonstrates strong compliance. QIR support is somewhat complete, with excellent export capabilities and limited import functionality. OpenQASM 2 support is comprehensive, though some  operations are unsupported. OpenQASM 3 support is not mentioned. Tket provides extensive format support through a variety of extension packages that cover Qiskit, Cirq, ProjectQ, PennyLane, Braket, and Quil. They actively participate in standardization efforts, including the OpenQASM Technical Steering Committee and QIR Alliance. Tket demonstrates excellent interoperability with other quantum frameworks through native extensions.
\textit{Qiskit} shows mixed compliance. QIR support is somewhat incomplete, limited to export-only through deprecated third-party tools. OpenQASM support is somewhat complete, with comprehensive OpenQASM 3 export, but the OpenQASM 2 exporter has known incompatibilities with its own importer, and OpenQASM 3 import remains experimental. Additional format support is somewhat limited, relying primarily on QPY serialization and OpenQASM bridges. They show active involvement in standardization, leading OpenQASM development and governance. Interoperability with other frameworks is moderate and achieved through third-party tools rather than native extensions.
\textit{ProjectQ} shows minimal compliance. QIR support is absent, and OpenQASM support is very limited. The framework does not support additional formats such as Quil. ProjectQ demonstrates no involvement in standardization efforts, with the project inactive on GitHub for more than a year. Interoperability with other frameworks is poor, relying on third-party tools like pytket-projectq for conversion.

\begin{table}[tbh]
\centering
\scriptsize
\begin{tabular}{p{8.8cm}|c|c|c}
\toprule
\textbf{Question} & \textbf{Tket} & \textbf{Qiskit} & \textbf{ProjectQ} \\
\midrule
How complete or incomplete is the compiler's support for QIR? & 4 & 2 & 1 \\
How complete or incomplete is the compiler's support for OpenQASM (import and export)? & 3 & 4 & 1 \\
How extensive or limited is the compiler's support for additional quantum-circuit description formats (e.g. Quil)? & 4 & 2 & 1 \\
How active or inactive is the compiler's involvement in quantum-computing standardization efforts? & 5 & 5 & 1 \\
How well or poorly does the compiler interoperate with other quantum frameworks (e.g., Cirq)? & 5 & 4 & 2 \\
\midrule
\textbf{Average} & 4.2 & 3.4 & 1.2 \\
\bottomrule
\end{tabular}
\caption{Questions and results for the Standardization Compliance dimension.}
\label{tab:met-standardization}
\end{table}

\textbf{Device-Agnostic Compiler Architecture:}
Table~\ref{tab:met-device-agnostic} shows that \textit{Tket} has a modular design, separating concerns into distinct components. Its hardware-agnostic design is evident in the reusable components across different hardware. 
Tket shows excellent quantum gate set adaptability, with support for diverse gate sets and easy extensibility. Hardware agnosticism is a core design principle, evident in the efforts to implement hardware-independent passes and custom hardware constraints. In our Sonar Qube evaluation, Tket received an overall score of 4, with a perfect score of 5 in security and in maintainability, and a 2 in reliability. 
\textit{Qiskit} features a highly modular design with clear separation of concerns, evident in its API reference which differentiates between circuit construction, quantum information, transpilation, and other components. While there is a focus on accommodating IBM machines, the framework is designed to support easy adaptation of its components across various hardware platforms. Qiskit is adaptable to diverse quantum gate sets and prioritizes hardware agnosticism as a core design principle throughout its system. When analyzed with Sonar Qube, Qiskit received an overall score of 2.3, with a perfect score of 5 in maintainability, but received a 1 in security and in reliability.
\textit{ProjectQ} performs well regarding its device-agnostic compiler architecture. It features a modular design, allowing for the definition of new \texttt{Engines} for the compilation process and new gates. The central components are designed for easy adaptation across different hardware. However, the framework is adaptable to diverse quantum gate sets, with easy definition of new gates and validity checks in the backend class. Hardware agnosticism is a core design principle, evident in the retargetable compiler structure with modular backend delivery. In the Sonar Qube evaluation, ProjectQ received an overall score of 2.67, with an excellent score of 5 in maintainability, a 2 in reliability and a 1 in security.

\begin{table}[tbh]
\centering
\scriptsize
\begin{tabular}{p{8.8cm}|c|c|c}
\toprule
\textbf{Question} & \textbf{Tket} & \textbf{Qiskit} & \textbf{ProjectQ} \\
\midrule
How modular or monolithic is the compiler's overall architecture? & 5 & 5 & 5 \\
How hardware-agnostic or hardware-specific are the compiler's components? & 5 & 5 & 5 \\
How adaptable or limited is the compiler to diverse quantum gate sets? & 5 & 5 & 5 \\
How high or low is the priority of hardware-agnostic design within the compiler? & 5 & 5 & 5 \\
What score did the compiler's codebase receive in the Sonar Qube evaluation? & 4 & 2.3 & 2.67 \\
\midrule
\textbf{Average} & 4.8 & 4.46 & 4.53 \\
\bottomrule
\end{tabular}
\caption{Questions and results for the Device-Agnostic Compiler Architecture dimension.}
\label{tab:met-device-agnostic}
\end{table}

\textbf{Community and Ecosystem Integration:}
As seen in Table~\ref{tab:met-community}, \textit{Tket} has excellent integration, receiving the highest rating in all subcategories. It is integrated with many major providers and has packages for various frameworks. It is also integrated with smaller providers, which is in large parts due to Tket's active community. Tket offers a module system that integrates with the \ac{pip}, enabling easy extensions. Custom backend distribution is accessible through GitHub and \ac{pip}. The GitHub community is very active, with frequent updates, multiple contributors, and significant engagement metrics (34 contributors, 291 stars, 55 forks, 209 releases, 1512 commits to main Tket repository).
\textit{Qiskit} is integrated with many major providers, including IBM, IonQ, AWT, Braket, Quantinuum, and Rigetti. The framework shows consistent growth in supported backends, with a dedicated community on GitHub actively developing new integrations. Qiskit offers well-designed extension capabilities, allowing users to extend various aspects such as transpiler passes and transpilers for new backends. Custom backend distribution is easy to do, with the ability to install custom backends through \ac{pip}, making it easily distributable and accessible. The community engagement is exceptionally high, with 634 contributors, over 6700 stars, and over 2700 forks on GitHub. The project is very active, with frequent updates and a dedicated community organization providing extensions.
\textit{ProjectQ} is integrated with many major providers and platforms, including AQT, Braket, Azure, IBM, and IonQ. The growth trajectory for integrations is unclear, with no obvious recent additions and notable inactivity of the project on GitHub for over a year. Extension capabilities are present, but not well-documented for users. Custom backend distribution is possible through \ac{pip}'s optional components. Community engagement is moderate, with updates coming from a small subset of the 32 contributors and most issues and pull requests created over a year ago. The project has seen significant interest, with 952 stars and 283 forks on GitHub.

\begin{table}[tbh]
\centering
\scriptsize
\begin{tabular}{p{8.8cm}|c|c|c}
\toprule
\textbf{Question} & \textbf{Tket} & \textbf{Qiskit} & \textbf{ProjectQ} \\
\midrule
How well or poorly integrated is the compiler with major and minor quantum-hardware providers? & 5 & 5 & 4 \\
How strong or weak has the growth been in the number of supported backends over time? & 5 & 5 & 2 \\
How extensive or limited are the compiler's extension capabilities (e.g., plugin system)? & 5 & 5 & 4 \\
How easy or difficult is it to distribute a custom backend with the compiler? & 5 & 5 & 2 \\
How active or inactive is the community surrounding the compiler (e.g., commits, contributors, discussions)? & 5 & 5 & 2 \\
\midrule
\textbf{Average} & 5 & 5 & 2.8 \\
\bottomrule
\end{tabular}
\caption{Questions and results for the Community and Ecosystem Integration dimension.}
\label{tab:met-community}
\end{table}

\subsection{User Study for Documentation and API Quality of Tket, Qiskit and ProjectQ}

In the following, we describe the execution and results of the user study.

\textbf{User Study Execution:}
For this study, we selected six post- and undergraduate students of computer science, all of which had at least basic knowledge of quantum software engineering. Some of the participants had prior experience with some of the evaluated compilers. To minimize setup overhead, each participant was instructed to use their own device to participate in the study. The participants used either MacOS or Windows. We provided a task sheet for each compiler. The task sheets are available at~\cite{zenodolink}. 
As a  concrete implementation task, they had to implement a backend for Stim~\cite{stimsimulator}, a simulator for quantum stabilizer circuits for quantum error correction. As this simulator is actively used and sufficiently complex, we determine that it represents a realistic backend implementation scenario. All participants were instructed to use the provided task sheet, without referring to the internet (e.g. search engines, forums) apart from links provided in the task sheet. These links included the documentation page for the compiler and the Stim simulator. In total, all participants invested around two working days into the implementation tasks.

Four participants (P1-P4) started by implementing the Tket backend, followed by Qiskit and ProjectQ. Two participants (P5-P6) started by implementing the Qiskit backend, followed by Tket and ProjectQ. As most of the participants had no experience with backend implementations, we chose to start with Tket and Qiskit rather than ProjectQ, as they have more comprehensive documentation.
Each participant implemented a backend for all three compilers and rated their experience in a questionnaire~\cite{zenodolink} composed of 14 questions.

\textbf{User Study Results:}
The questionnaire covered prior experience with quantum software engineering and each compiler (Q1-Q3), documentation quality and clarity (Q4-Q8), API usability (Q9), implementation difficulty (Q10), and the necessity of provided hints (Q11). We determined that Q1-Q3 do not add relevant insight to the compilers retargetability, which is why it is not included in the final retargetability score calculation.

Table~\ref{tab:usability} shows the detailed results of the questionnaire for all three compilers.
All participants had a low to moderate rate of experience in quantum software engineering in general (Q1) averaging 3.67. Prior experience with the compilers (Q2) was low for Tket and ProjectQ, with a higher value for Qiskit, although this was not prior backend implementation experience, as seen in the results for the third question. Prior backend implementation experience (Q3) was generally rated as very low. All these factors combined make for a similarly low rate of prior experience with implementing compiler backends, making the results comparable with little bias, other than that created through experience during the execution of the study.

For \textit{Tket}, the participants rated the compiler especially well in documentation quality and clarity (Q4-Q8). The API usability received a good score. The implementation difficulty received a moderate score of 3.33, while the participants regarded the provided hints as hardly necessary or not necessary. This indicates that Tket is reasonably accessible to developers. 
Across questions Q4-Q11, Tket received the best scores among the evaluated compilers in seven out of eight cases.
\textit{Qiskit} is the only evaluated compiler with which participants had previous experience. 
The scores indicate that participants 1-4 largely developed more familiarity with backend implementation concepts through their prior Tket implementation. Qiskit performed slightly worse than Tket in terms of documentation quality and clarity (Q4-8) and API usability (Q9). The implementation was rated less difficult (Q10) than Tket, which may be due to prior experience. The participants also rated the hints as more necessary than with Tket (Q11).
For \textit{ProjectQ}, the compiler specific experience (Q2) averaged 1, showing that all participants had no prior ProjectQ experience. Backend implementation experience (Q3) also averaged 1, suggesting that despite completing Tket and Qiskit implementations earlier in the study, participants felt unprepared for the different architecture of ProjectQ. This indicates that the skills learned from the implementation of the Tket and Qiskit backends did not transfer effectively to ProjectQ. 
The compiler performed significantly worse than Tket and Qiskit across all evaluated questions, with particular issues in documentation quality and implementation process.

\begin{table}[tbh]
  \centering
  \scriptsize
  \begin{tabular}{l|c|c|c|c|c|c|c|c|c|c|c|c|c}
    \toprule
      & & \textbf{Q1} & \textbf{Q2} & \textbf{Q3} & \textbf{Q4} & \textbf{Q5} & \textbf{Q6} & \textbf{Q7} & \textbf{Q8} & \textbf{Q9} & \textbf{Q10} & \textbf{Q11} & \textbf{Q4-11} \\
    \midrule
    \textbf{Tket} & \textbf{P1} & 3 & 1 & 1 & 4 & 4 & 3 & 3 & 3 & 4 & 3 & 5 & \\
    & \textbf{P2} & 5 & 1 & 1 & 5 & 5 & 4 & 5 & 5 & 4 & 4 & 2 & \\
    & \textbf{P3} & 3 & 1 & 1 & 5 & 4 & 4 & 5 & 5 & 4 & 3 & 4 & \\
    & \textbf{P4} & 5 & 1 & 1 & 4 & 4 & 5 & 4 & 5 & 4 & 2 & 3 & \\
    & \textbf{P5} & 2 & 1 & 1 & 5 & 5 & 5 & 5 & 5 & 5 & 5 & 5 & \\
    & \textbf{P6} & 4 & 2 & 1 & 5 & 5 & 5 & 5 & 5 & 4 & 3 & 5 & \\
    \cmidrule{2-14}
    & & 3.67 & 1.17 & 1 & 4.67 & 4.5 & 4.33 & 4.5 & 4.67 & 4.17 & 3.33 & 4 & \textbf{4.27} \\
    \midrule
    \textbf{Qiskit} & \textbf{P1} & 3 & 3 & 2 & 4 & 4 & 4 & 3 & 4 & 2 & 4 & 5 & \\
    & \textbf{P2} & 5 & 4 & 2 & 4 & 4 & 4 & 4 & 5 & 4 & 4 & 5 & \\
    & \textbf{P3} & 3 & 4 & 1 & 5 & 5 & 5 & 5 & 5 & 4 & 5 & 5 & \\
    & \textbf{P4} & 5 & 5 & 4 & 5 & 3 & 4 & 4 & 5 & 3 & 4 & 2 & \\
    & \textbf{P5} & 2 & 2 & 1 & 4 & 4 & 3 & 4 & 2 & 4 & 3 & 1 & \\
    & \textbf{P6} & 4 & 3 & 1 & 4 & 4 & 3 & 3 & 3 & 4 & 3 & 1 & \\
    \cmidrule{2-14}
     & & 3.67 & 3.5 & 1.83 & 4.33 & 4 & 3.83 & 3.83 & 4 & 3.5 & 3.83 & 3.17 & \textbf{3.81} \\
    \midrule
    \textbf{ProjectQ} & \textbf{P1} & 3 & 1 & 1 & 2 & 1 & 2 & 2 & 2 & 2 & 2 & 5 & \\
    & \textbf{P2} & 5 & 1 & 1 & 1 & 1 & 1 & 2 & 3 & 2 & 1 & 5 & \\
    & \textbf{P3} & 3 & 1 & 1 & 1 & 1 & 1 & 1 & 1 & 1 & 1 & 1 & \\
    & \textbf{P4} & 5 & 1 & 1 & 1 & 1 & 2 & 2 & 4 & 2 & 1 & 1 & \\
    & \textbf{P5} & 2 & 1 & 1 & 3 & 3 & 2 & 4 & 1 & 2 & 2 & 2 & \\
    & \textbf{P6} & 4 & 1 & 1 & 1 & 1 & 1 & 1 & 3 & 2 & 1 & 5 & \\
    \cmidrule{2-14}
     & & 3.67 & 1 & 1 & 1.5 & 1.33 & 1.5 & 2 & 2.33 & 1.83 & 1.33 & 3.17 & \textbf{1.88} \\
    \bottomrule
  \end{tabular}
  \caption{Documentation and API Quality study scores (higher is better for Q4-Q11)}
  \label{tab:usability}
\end{table}

\subsection{Overall Evaluation of Tket, Qiskit and ProjectQ}
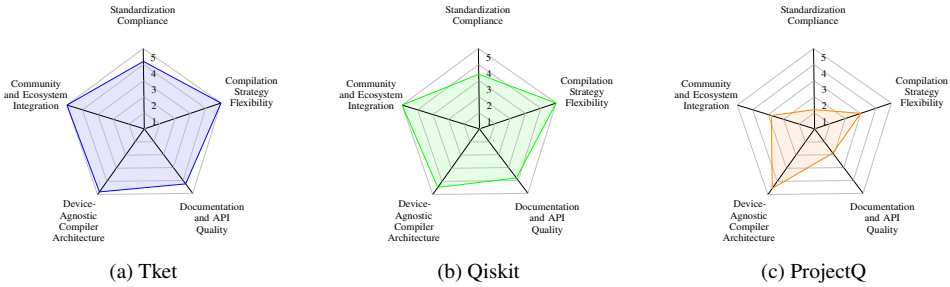
\begin{figure}[H]
     \centering
     \begin{subfigure}{0.3\textwidth}
         \centering
         \begin{adjustbox}{width=\textwidth}
         \begin{tikzpicture}[rotate=18.75]
\tkzKiviatDiagram[
        scale=0.5,
        space=0,
        label space=2,
        radial style/.style ={},
        radial  = 5,
        gap     = 1,  
        lattice = 5]{Compilation Strategy Flexibility,Standardization Compliance,Community and Ecosystem Integration,Device-Agnostic Compiler Architecture,Documentation and API Quality }
\tkzKiviatLine[thick,color=blue,mark=none,
               fill=blue!20,opacity=.5](5,4.2,5,4.8,4.27)   
\tkzKiviatGrad[prefix=,unity=1,suffix=](1)  
        \end{tikzpicture}
        \end{adjustbox}
    	\caption{Tket}
    	\label{fig:tket-spider}
     \end{subfigure}
     \hfill
     \begin{subfigure}{0.3\textwidth}
         \centering
         \begin{adjustbox}{width=\textwidth}
         \begin{tikzpicture}[rotate=18.75]
\tkzKiviatDiagram[
        scale=0.5,
        space=0,
        label space=2,
        radial style/.style ={},
        radial  = 5,
        gap     = 1,  
        lattice = 5]{Compilation Strategy Flexibility,Standardization Compliance,Community and Ecosystem Integration,Device-Agnostic Compiler Architecture,Documentation and API Quality }
\tkzKiviatLine[thick,color=green,mark=none,
               fill=green!20,opacity=.5](5,3.4,5,4.46,3.81)   
\tkzKiviatGrad[prefix=,unity=1,suffix=](1)  
        \end{tikzpicture}
        \end{adjustbox}
	\caption{Qiskit}
	\label{fig:qiskit-spider}
     \end{subfigure}
     \hfill
     \begin{subfigure}{0.3\textwidth}
         \centering
         \begin{adjustbox}{width=\textwidth}
         \begin{tikzpicture}[rotate=18.75]
\tkzKiviatDiagram[
        scale=0.5,
        space=0,
        label space=2,
        radial style/.style ={},
        radial  = 5,
        gap     = 1,  
        lattice = 5]{Compilation Strategy Flexibility,Standardization Compliance,Community and Ecosystem Integration,Device-Agnostic Compiler Architecture,Documentation and API Quality }
\tkzKiviatLine[thick,color=orange,mark=none,
               fill=orange!20,opacity=.5](3,1.2,2.8,4.53,1.88)   
\tkzKiviatGrad[prefix=,unity=1,suffix=](1)  
        \end{tikzpicture}
        \end{adjustbox}
	\caption{ProjectQ}
	\label{fig:projectq-spider}
     \end{subfigure}
        \caption{Retargetability scores for all dimensions represented as a spider diagram}
        \label{fig:spider-diagrams}
\end{figure}

By combining metric-based assessment and user study scores, as can be seen in Figure~\ref{fig:spider-diagrams}, we answer \textbf{RQ2}:
\textit{Tket} demonstrates the best retargetability, combining excellent performance across all metric categories with practical usability, as reflected in the best user study score of \textbf{4.65}. 
\textit{Qiskit} with a score of \textbf{4.33} demonstrates strong retargetability, but with more trade-offs compared to Tket. 
\textit{ProjectQ} demonstrates the worst retargetability among the three compilers with a score of \textbf{2.68}. 
While ProjectQ achieves good scores in device-agnostic architecture, it is lacking in practical usability due to a lack of documentation and compliance. 

\subsection{Threats to Validity}
In the paper, we proposed a metric that conveys many dimensions to assess retargetability. Although we have described a detailed reasoning for all dimensions, one could argue that those dimensions are not representative enough. Additionally, when calculating the retargetability score, all dimensions were given the same weights, which could easily be adjusted. 
Though best practices and state-of-the-art is lacking in this area, we are confident that we were able to oppose this threat by basing our reasoning on extensive literature research and our expert knowledge. Furthermore, we hope for further contributions in the field, improving our understanding of the subject and therefore improving future scoring attempts. 
The design of the survey instrument can influence the results obtained by the user study. To mitigate this threat, we considered best practices in survey design according to Dillman et al.~\cite{Dillman2024}.
As an assessment conducted by individuals is prone to subjective bias that is caused by personal opinions, varying experience levels or background, and other things. This work is no exception to this.
We oppose this threat by conducting the user study with as many participants as available.
Although a total number of 6 participants seems low, it is important to consider that participants with previous experience in the quantum software engineering domain that can also spare large time frames to implement experimental quantum backends are hard to obtain. We conclude that an increase of participants could improve the results of the study. Similar issues approach when discussing the selection of compilers. Especially the metric-based assessment and the design of the implantation task for the user study require advanced knowledge of the concrete compiler. It is possible to argue that an evaluation of more compilers would improve the quality of this study. However, we are constrained by the available resources that we have. A mitigation is to generalize this work through conducting additional studies and revisiting this topic in future work. 
\section{Conclusion}
In this paper, we proposed a methodology for measuring five dimensions of retargetability of quantum compilers: a four-dimensional metric to measure retargetability across the dimensions Compilation Strategy Flexibility, Standardization Compliance, Community and Ecosystem Integration and Device-Agnostic Compiler Architecture, and a user study that evaluates Documentation and API Quality.
We then evaluated the retargetability of Tket, Qiskit, and ProjectQ.
We observed that Tket performed best across all dimensions, closely followed by Qiskit. For ProjectQ we observed worse retargetability.
This research could be extended further by analyzing additional compilers, such as Weaver~\cite{kirmemics2025weaver}, which was released after the execution of this research and claims to be retargetable.

% \begin{acronym}[XXXXXX]
%     \acro{tva}[TVA]{Institute for Test, Validation and Analysis of Software Intensive Systems}
%     \acro{kit}[KIT]{Karlsruhe Institute of Technology}
%     \acro{kastel}[KASTEL]{Institute of Information Security and Dependability}
%     \acro{rsa}[RSA]{Rivest–Shamir–Adleman}
%     \acro{nisq}[NISQ]{Noisy Intermediate-Scale Quantum}
%     \acro{sdk}[SDK]{Software Development Kit}
%     \acro{qir}[QIR]{Quantum Intermediate Representation}
%     \acro{openqasm}[OpenQASM]{Open Quantum Assembly Language}
%     \acro{pip}[pip]{Package Installer for Python}
% \end{acronym}

\section*{Acknowledgements}
This work has been partially supported by the German Federal Ministry of Research, Technology and Space in project FullStaQD under Grant No.: 01MQ25001F.

\printbibliography

@misc{IBM:QISKIT,
      title={Quantum computing with {Q}iskit},
      author={Javadi-Abhari, Ali and Treinish, Matthew and Krsulich, Kevin and Wood, Christopher J. and Lishman, Jake and Gacon, Julien and Martiel, Simon and Nation, Paul D. and Bishop, Lev S. and Cross, Andrew W. and Johnson, Blake R. and Gambetta, Jay M.},
      year={2024},
      doi={10.48550/arXiv.2405.08810},
      eprint={2405.08810},
      archivePrefix={arXiv},
      primaryClass={quant-ph}
}

@software{GOOGLE:CIRQ,
author = {{Cirq Developers}},
license = {Apache-2.0},
month = aug,
title = {{Cirq}},
url = {https://github.com/quantumlib/Cirq},
version = {1.6.1},
year = {2025}
}

@article{Sivarajah_2020,
   title={t|ket⟩: a retargetable compiler for NISQ devices},
   volume={6},
   ISSN={2058-9565},
   url={http://dx.doi.org/10.1088/2058-9565/ab8e92},
   DOI={10.1088/2058-9565/ab8e92},
   number={1},
   journal={Quantum Science and Technology},
   publisher={IOP Publishing},
   author={Sivarajah, Seyon and Dilkes, Silas and Cowtan, Alexander and Simmons, Will and Edgington, Alec and Duncan, Ross},
   year={2020},
   month=nov, pages={014003} 
}

@inproceedings{Salm2021_CompilerComparison,
author = {Salm, Marie and Barzen, Johanna and Leymann, Frank and
Weder, Benjamin and Wild, Karoline},
title = {{Automating the Comparison of Quantum Compilers
for Quantum Circuits}},
booktitle = {Proceedings of the 15th Symposium and Summer School on
Service-Oriented Computing (SummerSOC 2021)},
pages = {64--80},
publisher = {Springer International Publishing},
month = sep,
year = 2021,
doi = {10.1007/978-3-030-87568-8_4}
}

@article{Steiger_2018,
   title={ProjectQ: an open source software framework for quantum computing},
   volume={2},
   ISSN={2521-327X},
   url={http://dx.doi.org/10.22331/q-2018-01-31-49},
   DOI={10.22331/q-2018-01-31-49},
   journal={Quantum},
   publisher={Verein zur Forderung des Open Access Publizierens in den Quantenwissenschaften},
   author={Steiger, Damian S. and Häner, Thomas and Troyer, Matthias},
   year={2018},
   month=jan, 
   pages={49} 
}

@mastersthesis{Gierisch:AnaEvalQC,
  author = {Vincent Gierisch},
  title = {{Analysis and Evaluation of Quantum Compilers}},
  school = {Ostbayerische Technische Hochschule Regensburg},
  type = {Master Thesis},
  address = {Germany},
  month = MAY,
  year = 2024,
  note={Available online at \url{https://www.lfdr.de/Theses/2024/MA-Gierisch.pdf}, visited on October 28, 2024}
}

@article{quetschlich2023mqt,
  title={MQT Bench: Benchmarking software and design automation tools for quantum computing},
  author={Quetschlich, Nils and Burgholzer, Lukas and Wille, Robert},
  journal={Quantum},
  volume={7},
  pages={1062},
  year={2023},
  publisher={Verein zur F{\"o}rderung des Open Access Publizierens in den Quantenwissenschaften}
}

@article{nation2024benchmarking,
  title={Benchmarking the performance of quantum computing software},
  author={Nation, Paul D and Saki, Abdullah Ash and Brandhofer, Sebastian and Bello, Luciano and Garion, Shelly and Treinish, Matthew and Javadi-Abhari, Ali},
  journal={arXiv preprint arXiv:2409.08844},
  year={2024}
}

@article{kharkov2022arline,
  title={Arline benchmarks: Automated benchmarking platform for quantum compilers},
  author={Kharkov, Y and Ivanova, Alexandra and Mikhantiev, E and Kotelnikov, A},
  journal={arXiv preprint arXiv:2202.14025},
  year={2022}
}

@article{meijer2025comparison,
  title={A comparison of quantum compilers using a DAG-based or phase polynomial-based intermediate representation},
  author={Meijer-van de Griend, Arianne},
  journal={Journal of Systems and Software},
  volume={221},
  pages={112224},
  year={2025},
  publisher={Elsevier}
}

@inproceedings{Svore_2018, 
   series={RWDSL2018},
   title={{Q\#: Enabling Scalable Quantum Computing and Development with a High-level DSL}},
   url={http://dx.doi.org/10.1145/3183895.3183901},
   DOI={10.1145/3183895.3183901},
   booktitle={Proceedings of the Real World Domain Specific Languages Workshop 2018},
   publisher={ACM},
   author={Svore, Krysta and Geller, Alan and Troyer, Matthias and Azariah, John and Granade, Christopher and Heim, Bettina and Kliuchnikov, Vadym and Mykhailova, Mariia and Paz, Andres and Roetteler, Martin},
   year={2018},
   month=feb, collection={RWDSL2018}
}

@inproceedings{kissinger2020Pyzx,
    author = {Kissinger, Aleks and van de Wetering, John},
    title = {{PyZX: Large Scale Automated Diagrammatic Reasoning}},
    year = {2020},
    booktitle = {{\rm Proceedings 16th International Conference on} Quantum Physics and Logic, {\rm Chapman University, Orange, CA, USA., 10-14 June 2019}},
    editor = {Coecke, Bob and Leifer, Matthew},
    series = {Electronic Proceedings in Theoretical Computer Science},
    volume = {318},
    pages = {229-241},
    publisher = {Open Publishing Association},
    doi = {10.4204/EPTCS.318.14}
}

@article{amy2020staq,
  title={staq—A full-stack quantum processing toolkit},
  author={Amy, Matthew and Gheorghiu, Vlad},
  journal={Quantum Science and Technology},
  volume={5},
  number={3},
  pages={034016},
  year={2020},
  publisher={IOP Publishing}
}

@article{seidel2024qrisp,
  title={Qrisp: A framework for compilable high-level programming of gate-based quantum computers},
  author={Seidel, Raphael and Bock, Sebastian and Zander, Ren{\'e} and Petri{\v{c}}, Matic and Steinmann, Niklas and Tcholtchev, Nikolay and Hauswirth, Manfred},
  journal={arXiv preprint arXiv:2406.14792},
  year={2024}
}

@article{li2023qasmbench,
  title={Qasmbench: A low-level quantum benchmark suite for nisq evaluation and simulation},
  author={Li, Ang and Stein, Samuel and Krishnamoorthy, Sriram and Ang, James},
  journal={ACM Transactions on Quantum Computing},
  volume={4},
  number={2},
  pages={1--26},
  year={2023},
  publisher={ACM New York, NY}
}

@inproceedings{kirmemics2025weaver,
  title={Weaver: A Retargetable Compiler Framework for FPQA Quantum Architectures},
  author={K{\i}rmemi{\c{s}}, O{\u{g}}uzcan and Rom{\~a}o, Francisco and Giortamis, Emmanouil and Bhatotia, Pramod},
  booktitle={Proceedings of the 23rd ACM/IEEE International Symposium on Code Generation and Optimization},
  pages={299--316},
  year={2025}
}

@inproceedings{dangwal2024compass,
  title={COMPASS: Compiler Pass Selection For Improving Fidelity Of NISQ Applications},
  author={Dangwal, Siddharth and Ravi, Gokul Subramanian and Seifert, Lennart Maximilian and Das, Poulami and Sud, James and Chong, Frederic T},
  booktitle={2024 IEEE International Conference on Rebooting Computing (ICRC)},
  pages={1--14},
  year={2024},
  organization={IEEE}
}

@book{Dillman2024,
author = {Dillman, Don A. and Smyth, Jolene D. and Christian, Leah Melani},
title = {Internet, Phone, Mail, and Mixed-Mode Surveys: The Tailored Design Method},
year = {2014},
isbn = {1118456149},
publisher = {Wiley Publishing},
edition = {4th}
}

@BOOK{national2018quantum,
  author    = {{National Academies of Sciences, Engineering, and Medicine}},
  editor    = "Emily Grumbling and Mark Horowitz",
  title     = "Quantum Computing: Progress and Prospects",
  isbn      = "978-0-309-47969-1",
  doi       = "10.17226/25196",
  url       = "https://nap.nationalacademies.org/catalog/25196/quantum-computing-progress-and-prospects",
  year      = 2019,
  publisher = "The National Academies Press",
  address   = "Washington, DC"
}

@misc{compreviewqiu,
      title={A Comparative Review of Quantum Bits: Superconducting, Topological, Spin, and Emerging Qubit Technologies}, 
      author={Akash Chohan},
      month = NOV,
      year={2024},
      url={https://papers.ssrn.com/sol3/papers.cfm?abstract_id=4979773}
}

@misc{stimsimulator,
      title={Stim: a fast stabilizer circuit simulator}, 
      author={Craig Gidney},
      month = JUL,
      year={2021},
      url={https://quantum-journal.org/papers/q-2021-07-06-497/#}
}

@misc{sonarqube,
      title={SonarQube - Code quality and security}, 
      author={SonarSource},
      url={https://www.sonarsource.com/products/sonarqube/},
      note = {Accessed: 2025-12-01}
}

@article{preskill2018quantum,
  title={Quantum computing in the NISQ era and beyond},
  author={Preskill, John},
  journal={Quantum},
  volume={2},
  pages={79},
  year={2018},
  publisher={Verein zur F{\"o}rderung des Open Access Publizierens in den Quantenwissenschaften}
}

@misc{murilloChallengesQuantumSoftware2024,
  title = {Challenges of {{Quantum Software Engineering}} for the {{Next Decade}}: {{The Road Ahead}}},
  shorttitle = {Challenges of {{Quantum Software Engineering}} for the {{Next Decade}}},
  author = {Murillo, Juan M. and {Garcia-Alonso}, Jose and Moguel, Enrique and Barzen, Johanna and Leymann, Frank and Ali, Shaukat and Yue, Tao and Arcaini, Paolo and Castillo, Ricardo P{\'e}rez and {de Guzm{\'a}n}, Ignacio Garc{\'i}a Rodr{\'i}guez and Piattini, Mario and {Ruiz-Cort{\'e}s}, Antonio and Brogi, Antonio and Zhao, Jianjun and Miranskyy, Andriy and Wimmer, Manuel},
  year = {2024},
  month = apr,
  number = {arXiv:2404.06825},
  eprint = {2404.06825},
  primaryclass = {cs},
  publisher = {arXiv},
  urldate = {2024-07-30},
  archiveprefix = {arXiv},
  langid = {english}
}

@misc{zenodolink,
  author       = {Southall, Luke and
                  Ammermann, Joshua and
                  Kelmendi, Rinor and
                  Eichhorn, Domenik and
                  Schaefer, Ina},
  title        = {Methodology for Retargetability Assessment of
                   Quantum Compilers
                  },
  month        = dec,
  year         = 2025,
  publisher    = {Zenodo},
  doi          = {10.5281/zenodo.17780085},
  url          = {https://doi.org/10.5281/zenodo.17780085},
}

\end{document}